# Hybridization-Controlled Pseudogap State in the Quantum Critical Superconductor CeCoIn$_5$

Harim Jang[1], Vuong Thi Anh Hong[1], Jihyun Kim[1], Xin Lu[2], and Tuson Park[1,3,*]

[1] *Department of Physics, Sungkyunkwan University, Suwon 16419, Republic of Korea*

[2] *Center for Correlated Matter and Department of Physics, Zhejiang University, Hangzhou 310058, China*

[3] *Center for Quantum Materials and Superconductivity (CQMS), Sungkyunkwan University, Suwon 16419, Republic of Korea*

**The origin of the partial suppression of the electronic density states in the enigmatic pseudogap behavior, which is at the core of understanding high-$T_c$ superconductivity, has been hotly contested as either a hallmark of preformed Cooper pairs or an incipient order of competing interactions nearby. Here, we report the quasi-particle scattering spectroscopy of the quantum critical superconductor CeCoIn$_5$, where a pseudogap with energy $\Delta_g$ was manifested as a dip in the differential conductance ($dI/dV$) below the characteristic temperature of $T_g$. When subjected to external pressure, $T_g$ and $\Delta_g$ gradually increase, following the trend of increase in quantum entangled hybridization between Ce 4$f$ moment and conduction electrons. On the other hand, the superconducting (SC) energy gap and its phase transition temperature shows a maximum, revealing a dome shape under pressure. The disparate dependence on pressure between the two quantum states shows that the pseudogap is less likely involved in the formation of SC Cooper pairs, but rather is controlled by Kondo hybridization, indicating that a novel type of pseudogap is realized in CeCoIn$_5$.**

*tp8701@skku.edu

The pseudogap, reminiscent of the energy gap from the depression of the density of states or a partial gap near the Fermi level, is of great interest because of its potential connection to the intertwined quantum states emergent in strongly correlated systems [1-8]. Attractive, yet puzzling, is the observation of the pseudogap in various unconventional superconductors, indicating that understanding its microscopic origin is crucial to resolving the mysteries of high-temperature superconductivity [4,9-13]. Although extensive studies have been conducted to elucidate the origin of the pseudogap, its relationship with unconventional superconductivity is still controversial owing to the disorder inherent in chemical substitution, close proximity to other intertwined quantum states, and large thermal fluctuations [3,6,7,14].

In the prototypical unconventional superconductor $CeCoIn_5$, which is located near a quantum critical point in its stoichiometric form, a pseudogap feature in the normal state above the superconducting transition temperature ($T_c$) has been reported from spectroscopic and transport measurements [9,15-17]. The precursor state of Cooper pairs was proposed as the origin of a pseudogap from scanning tunneling spectroscopy (STS) measurements, where depletion in the differential conductance ($dI/dV$) near the Fermi level was observed below 3.0 K with an energy scale of 1 meV [15]. Deviating from the $T$-linear non-Fermi liquid behavior, the electrical resistivity was suppressed below 3 K, which was ascribed to a decrease in the scattering rate owing to the preformed Cooper pairs [16]. In contrast, the close proximity to the antiferromagnetic quantum critical point (AFM QCP) of $CeCoIn_5$ suggests that a new form of competing orders may be realized in the normal state, requiring further study on the nature of the pseudogap in $CeCoIn_5$ [7,16,18].

Here, we report the hybridization-controlled novel pseudogap state in the quantum critical superconductor $CeCoIn_5$. A dip overlaid with the Fano-resonance line in differential conductance $dI/dV$ is observed via quasi-particle scattering spectroscopy (QSS) at temperatures higher than $T_c$, indicating the appearance of a pseudogap with energy $\Delta_g$ in the normal state. In stark contrast to the dome-shape pressure dependence of the superconducting (SC) transition temperature, $\Delta_g$ and its emerging temperature ($T_g$) monotonically increase under pressure and follow the trend of the transport Kondo coherence temperature ($T_{coh}$). The Fano-resonance asymmetry in $dI/dV$ under pressure increases with decreasing temperature and reveals a scaling behavior against the reduced temperature ($T/T_{sat}$) in the high-temperature regime, signifying that $T_{sat}$ is the onset of Kondo lattice coherence. With a further decrease in temperature, it deviates from the scaling and shows a peak or kink at $T_{max}$. When plotted together with pressure, the characteristic temperatures of $T_g$, $T_{max}$, and $T_{coh}$ follow the same pressure dependence with the pseudogap energy $\Delta_g$, but is different from that of $T_c$. These discoveries reveal that the nature of the pseudogap in CeCoIn$_5$ is not related to formation of Cooper electron pairs, but is pertinent to Kondo hybridization.

Figure 1(a) shows the generic temperature–pressure phase diagram of CeCoIn$_5$, where $T_c$ peaked at 1.5 GPa shows a dome shape and the transport Kondo coherence temperature determined from the resistivity maximum monotonically increases under pressure [16,19]. The dashed line is the projected pseudogap temperature based on a preformed Cooper pair scenario, which decreases under pressure and disappears near the end of the SC dome [16]. The inset shows a schematic view of the point-contact junction surrounded by a pressure medium; see the Supplemental Material (SM) for the details [20]. In QSS obtained from the point-contact technique, two distinct paths of the transmitting electrons to heavy-fermion metal are considered: tunneling into the conduction band (*c*-channel) and the heavy-electron band (*f*-channel) [21-23]. Quantum interference between these two paths produces the Fano line in $dI/dV$ with a peak centered at the Kondo resonance energy ($\varepsilon$), which can be described by the generalized Fano resonance model for the Kondo lattice [21]:

$$\frac{dI}{dV}(V,T) = g_0 + g_I \int \frac{df(E-V,T)}{dV} \frac{|q-\tilde{E}|^2}{1+\tilde{E}^2} dE \qquad (1)$$

The term $|q - \tilde{E}|^2/1 + \tilde{E}^2$ represents the simple Fano line, where $\tilde{E} = (E - \varepsilon)/[(aV)^2 + \gamma^2]^{1/2}$ is a phenomenologically modified expression. The Fano parameter $q = q_1 + iq_2$ reflects the tunneling ratio between the two paths. Here, $q_2$ describes the direct tunneling into the $c$–$f$ coupled state around $\varepsilon$ without interference, which could lead to the destruction of the Fano interference. The variables $f(E,T)$, $g_0$, and $\gamma$ are the Fermi function, background conductance, and scattering rate at the zero-bias voltage within the $f$-channel, respectively (see section C in the SM for more details [20]).

The temperature evolution of $dI/dV$ for CeCoIn$_5$ at ambient pressure is selectively displayed from 70 K (top) to 1.5 K (bottom) with a rigid offset for clarity in Fig. 1(b). The two-peak structure with spacing of ~2.4 meV at 1.5 K is ascribed to the Andreev reflection (AR), which is the hallmark of superconductivity [24]. At temperatures above $T_c$ of 2.3 K, the AR is completely suppressed, whereas a dip feature in $dI/dV$ is observed near zero-bias voltage, as indicated by the arrow at 3.0 K. The solid lines are the best fit of the Fano resonance model, showing good agreement with the experimental data (see Fig. S3, S7, and S8 in the SM for the detailed analysis for the full range of pressure and temperature [20]). The Kondo resonance peak near $\varepsilon = 3.05$ meV is gradually suppressed as the temperature increases, and is hardly noticeable at temperatures above $T_{coh} \sim 45$ K, which is consistent with previous results [21,25]. In contrast, the asymmetry in $dI/dV$ still holds even at 70 K.

The differential conductance of CeCoIn$_5$ divided by the Fano contribution, $dI/dV$ / $dI/dV|_{FANO}$, is plotted for several temperatures with an offset for clarity in Fig. 2(a), where the evolution of the dip feature as a function of temperature is highlighted by the color shades (see Fig. S4 in the SM for raw data [20]). With increasing temperature, the amplitude of the dip is suppressed and completely smeared out above the characteristic temperature of $T_g$, whereas the width of the dip ($\Delta_g$) does not decrease as observed in a normal order parameter. The magnetic field dependence of the dip feature in $dI/dV$ of CeCoIn$_5$ at 2.2 K and 3.0 K at ambient pressure is summarized in Fig. 2(b) and 2(c), respectively. At 2.2 K, which is slightly below $T_c$, the AR peak is observed at 0 T and is suppressed under a magnetic field. When the field is larger than the upper critical field, the AR is completely suppressed, but a dip feature is introduced instead. However, at 3.0 K, which is slightly higher than $T_c$, a dip feature appears, while AR peak is absent. The dip is gradually suppressed with the increasing field and is not apparent above 4.0 T. The energy width $\Delta_g$ (~ 0.8 meV), which is determined as the half of the full width at half maximum (FWHM) at just above $T_c$, is comparable to that of the pseudogap reported in STM on CeCoIn$_5$ [9,15,17]. As increasing temperature, the dip was filled up smoothly and almost independent of temperature, as observed in numerous high-$T_c$ cuprates [26]. These features suggest that the dip in $dI/dV$ arises from the opening of the pseudogap of CeCoIn$_5$ near the Fermi level.

The dependence on temperature at 0 T and magnetic field at 3.0 K of the normalized differential conductance, $dI/dV$ / $dI/dV|_{FANO}$, are depicted as a color contour plot for representative pressures in Fig. 3(a) – (e) and Fig. 3(f) – (j), respectively. The plume shape within the energy gap $\Delta_g$ near the zero-bias voltage describes the pseudogap feature, where the pseudogap regime is enlarged with increasing pressure on both the temperature and magnetic field vs. bias voltage planes. The onset temperature $T_g$ is determined using the criteria of –0.2 % boundary in *dI/dV* / *dI/dV*|FANO for all pressure ranges (see Fig. S4 in the SM for more details [20]). The application of pressure changes the gap feature systematically: $T_g$ of 5.5(±1.0) K at 0.00 GPa increases to 12.5(±2.1) K at 2.34 GPa, whereas $2\Delta_g$ of 1.60(±0.21) meV at 0.00 GPa increases to 3.96(±0.71) meV at 2.34 GPa. In addition, the critical field required to destroy the pseudogap feature at 3.0 K increases from 3.0 T at 0.0 GPa to a field higher than 8.0 T at 2.34 GPa (see Fig. S9 and S10 in the SM for details [20]). The suppression of the gap amplitude under a magnetic field, as shown in Fig. 3, is different from that of temperature, implying that the magnetic field not only fills the gapped feature, but also suppresses the onset temperature of pseudogap feature.

The degree of Fano asymmetry in *dI/dV* can be defined by the ratio of the differential conductance value at a specific energy of positive and negative bias [25]. We select 10 mV as a criterion to define the degree of asymmetry, i.e. $A$ (%) = 100 × (d*I*/d*V*|$_{-10mV}$ − d*I*/d*V*|$_{+10mV}$) / (d*I*/d*V*|$_{+10mV}$) (see Fig. S10 in the SM for the results with different criteria [20]). Because Ce 4*f* electrons are localized at high temperatures, the *c*–*f* coupling strength is expected to be suppressed with increasing temperature, resulting in a decrease in $A$ and saturate to zero above the characteristic temperature, $T_{sat}$. The temperature dependence of $A(T)$ follows the semi-phenomenological relation of the two-fluid model, depicted as a dashed line in Fig 4(a) in the high-temperature regime [27]:

$$A(T) \propto \left(1 - \frac{T}{T_{sat}}\right)^{3/2} \left[1 - \ln\left(\frac{T}{T_{sat}}\right)\right] \qquad (2)$$

Here, the first term is an order parameter that characterizes the collective coherent phenomenon, while the second term is the average effective mass of heavy quasiparticle that increases logarithmically. At ambient pressure, the asymptotic temperature $T_{sat}$ is approximately 150 K, which is much higher than the transport coherence temperature $T_{coh} \sim 45$ K [27-29], but comparable with the onset temperature of the collective $c$–$f$ hybridization gap reported by angle-resolved photoemission spectroscopy (ARPES) and ultrafast optical spectroscopy [30,31]. In the low-$T$ regime far below $T_{sat}$, $A(T)$ starts to deviate from the two-fluid description and shows a maximum (or kink) at the characteristic temperature $T_{max}$, indicated by the arrows in Fig 4(a). We note that such discrepancy between energy-resolved spectroscopy and transport measurement on $T_{sat}$ was also reported in other Kondo lattice YbRh$_2$Si$_2$ [32].

When subjected to external pressure, the magnitude of Fano asymmetry $A$ decreases in the low-$T$ regime, but collapses onto a single curve in the high-$T$ regime as a function of the reduced temperature of $T/T_{sat}$ for different pressures. As shown in the inset of Fig 4(a), both $T_{sat}$ and $T_{max}$ linearly increase with pressure; this can be ascribed to the enhanced $c$–$f$ coupling strength. The maximum Fano asymmetry at $T_{max}$ may be used as a possible indicator of the complete transition in a broad crossover from the Kondo incoherent to the entangled quantum coherent state. Supporting this interpretation, ARPES measurements show that the saturating behavior of the $f$-electron weight occurs near $T_{max}$ (~20 K) in CeCoIn$_5$ [30,31].

Figure 4(b) summarizes the comprehensive $P$–$T$ phase diagram of CeCoIn$_5$ overlaid with the normalized zero-bias-conductance (ZBC) color contour plot, where the ZBC is normalized to its value at $T_c$ for clarity. The characteristic temperatures of $T_g$, $T_{coh}$, $T_{max}$, $T_{c,\rho}$, and $T_{c,QSS}$ are plotted on the left ordinate. Here, $T_{coh}$ and $T_{max}$ are multiplied by a factor of 0.1 and 0.3, respectively, for comparison with other temperature scales. The width of the suppressed $dI/dV$, which is attributed to the opening of a pseudogap $2\Delta_g$, is plotted on the right ordinate. All the temperature and energy scales that characterize the low-energy physics in the normal state nearly triple as the pressure increases from 1 bar to 2.34 GPa and follow the linear-in-$P$ trend (see the dashed line). The pressure dependences of both $T_g$ and $\Delta_g$, which characterize the pseudogap, deviate significantly from that of $T_c$, implying that the pseudogap in CeCoIn$_5$ is not involved in the Cooper pair formation process. Instead, the pseudogap temperature and its amplitude closely follow the pressure dependence of $T_{coh}$ (see Fig. S13 in the SM [20]), suggesting that the depletion of the density of states in the normal state originates from the collective behavior of $c$–$f$ hybridization.

In high-$T_c$ cuprates, a pseudogap determined from the depletion in the quasiparticle density of states below $T^*$ ($\gg T_c$) has almost a temperature-independent energy width, and the SC gap-to-$T^*$ ratio ($2\Delta_{SC}/k_B T^*$) is approximately 4.3 for various members of high-$T_c$ cuprates [26]. Similarly, the pseudogap of CeCoIn$_5$ is distinct from the temperature response of the ordinary order parameter and shows filling-up behavior with increasing temperature (see Fig. 3 and Fig. S5 in the SM for more details [20]). The SC gap-to-$T_g$ ratio of CeCoIn$_5$, contrarily, is approximately 2.6($\pm$0.98) at ambient pressure, which is substantially smaller than those of high-$T_c$ cuprates. Furthermore, the response to doping concentration (or pressure) is disparate between the two families of unconventional superconductors. With increasing doping level, the onset temperature as well as the size of the pseudogap in the cuprates decrease and disappear at the critical concentration above which the SC phase is suppressed [4,26]. In CeCoIn$_5$, in striking contrast, the pseudogap behavior continuously increases even in the high-pressure regime, where superconductivity is suppressed. These contrasting behaviors against the SC transition temperature indicate that the origin of the pseudogap of CeCoIn$_5$ is different from that of the high-$T_c$ cuprates, emphasizing the importance of more comprehensive tests to elucidate the nature and mechanism of pseudogap phenomena in strongly correlated superconductors.

A few prior spectroscopic studies on CeCoIn$_5$ surface have observed the smooth evolution of SC gap with lowering temperature, suggesting preformed Cooper pairs above $T_c$ [15,33]. Since both the SC gap and pseudogap feature is revealed as a dip structure in the *dI/dV* probed by tunneling spectroscopic method, it is difficult to determine if the pseudogap behavior is intertwined with the superconductivity below bulk $T_c$ or continuously evolved into SC gap. Even though such a smooth *dI/dV* evolution across $T_c$ also has been widely observed in high-$T_c$ cuprates, the origin of pseudogap still awaits an answer [26]. Taken together with a theoretical explanation that pseudogap behavior in STS of CeCoIn$_5$ can be explained using an electronic structure of heavy quasiparticle, the pseudogap in CeCoIn$_5$ is also controversial [34]. We finally note that several recent STS studies emphasized the importance of *c–f* entanglement on the pseudogap development in CeCoIn$_5$, as is our study (see Fig. S6 in the SM [20]) [9,17].

In summary, the *f* electron delocalization and the nature of pseudogap in CeCoIn$_5$ have been studied via energy-resolved spectroscopy and systematic pressure control. The evolution of *dI/dV* reveals the Fano asymmetry even at temperatures much higher than the transport Kondo coherence temperature, $T_{coh}$. Signifying the onset of itinerant heavy quasiparticles at $T_{sat}$, the Fano asymmetry under pressure shows a scaling behavior as a function of the reduced temperature of $T/T_{sat}$ in the high-$T$ regime. With decreasing temperature, a dip feature in *dI/dV* is observed around the zero-bias voltage in the normal state owing to the opening of the pseudogap. Applied pressure gradually increases both the pseudogap width $\Delta_g$ and onset temperature $T_g$, which is opposite of the pressure dependence of the SC gap, but is similar to that of $T_{coh}$. These discoveries underline that the origin of the pseudogap in CeCoIn$_5$ is not related to the formation of SC Cooper pairs, but arises from the collective Kondo hybridization effects. The successful demonstration of the energy-dependent transport spectroscopy under pressure is not only pertinent to the analysis of the pseudogap in the quantum critical compound

CeCoIn$_5$, but is also expected to play a pivotal role in elucidating the origin of pseudogap and their connection to superconductivity in other families of unconventional superconductors, such as high-$T_c$ cuprates.

## ACKNOWLEDGEMENTS

We acknowledge fruitful discussion with J. D. Thompson and W. K. Park. This work was supported by the National Research Foundation (NRF) of Korea through a grant funded by the Korean Ministry of Science and ICT (No. 2021R1A2C2010925). Work at Zhejiang University was supported by the National Key Research and Development Program of China (No. 2017YFA0303101).

**Figure Captions**

**FIG. 1.** (a) Generic $P$–$T$ phase diagram of CeCoIn$_5$. The green dashed line is a pseudogap boundary predicted from previous works [16]. The inset shows the schematic view for the quasi-particle scattering spectroscopy under pressure, where the junction is made on the $a$–$b$ crystallographic plane. (b) Differential conductance ($dI/dV$) of CeCoIn$_5$ at representative temperatures and at ambient pressure. The solid lines are the best fits using the Fano resonance model (see the main text for details). The arrow at 3.0 K indicates a dip feature near the zero-bias voltage. All the curves are rigidly shifted for clarity.

**FIG. 2.** (a) Temperature-dependent normalized differential conductance, $dI/dV$ / $dI/dV|_{FANO}$, at ambient pressure, where $dI/dV|_{FANO}$ is the best fit of the Fano resonance model. The dependences on the magnetic field of $dI/dV$ / $dI/dV|_{FANO}$ at 2.2 K and 3.0 K are plotted in (b) and (c), respectively. The shaded area highlights the minimum near the zero-bias voltage. For clarity, all the curves are rigidly shifted against the bottom curves of 2.2 K, 0 T, and 0 T for panel (a), (b), and (c), respectively. In (c), the spectra at 0 T and 1 T are multiplied by a factor of 0.5 for comparison.

**FIG. 3.** Color contour plot of $dI/dV$ / $dI/dV|_{FANO}$ at (a) $P$ = 0.00 GPa, (b) 1.00 GPa, (c) 1.52 GPa, (d) 1.80 GPa, and (e) 2.34 GPa in the temperature ($T$) vs bias-voltage ($V$) axes. Color contour plot of $dI/dV$ / $dI/dV|_{FANO}$ under magnetic field at 3.0 K and at (f) $P$ = 0.00 GPa, (g) 1.00 GPa, (h) 1.52 GPa, (i) 1.80 GPa, and (j) 2.34 GPa in the field ($B$) vs bias-voltage ($V$) axes. Dark brown and blue in the color scale describe positive values, whereas green represents

negatives. All the contour plots are shown with the same scale for comparison and the dashed lines are guides to the eye.

**FIG. 4.** (a) Fano asymmetry, determined at 10 mV, as a function of reduced temperature, $T/T_{\text{sat}}$. The dashed line is the universal temperature dependence curve estimated from the two-fluid model (see the main text for details). The arrows indicate $T_{\max}$, where $A(T)$ has a maximum or kink. Pressure-dependent $T_{\text{sat}}$ and $T_{\max}$ are shown in the inset, where the dashed line is a guide to the eye. (b) Comprehensive $P$–$T$ phase diagram overlaid with the color contour plot of the normalized zero-bias-conductance, (ZBC) / (ZBC at $T_c$). $T_g$, $T_{\text{coh}}$, $T_{\max}$, $T_{c,\rho}$, and $T_{c,\text{QSS}}$, plotted on the left ordinate, stand for characteristic temperatures for pseudogap, transport Kondo coherence, maximum Fano asymmetry, and SC transition determined from resistivity and QSS, respectively, where $T_{\text{coh}}$ and $T_{\max}$ are consecutively multiplied by a factor of 0.1 and 0.3 for comparison (see Fig. S14 in the SM for more details [20]). The pseudogap width $2\Delta_g$ is plotted on the right ordinate. The dashed lines are visual guides. Errors represent uncertainty in determining the characteristic temperatures as well as the pseudogap based on the standard deviation of each definition.

**Figures**

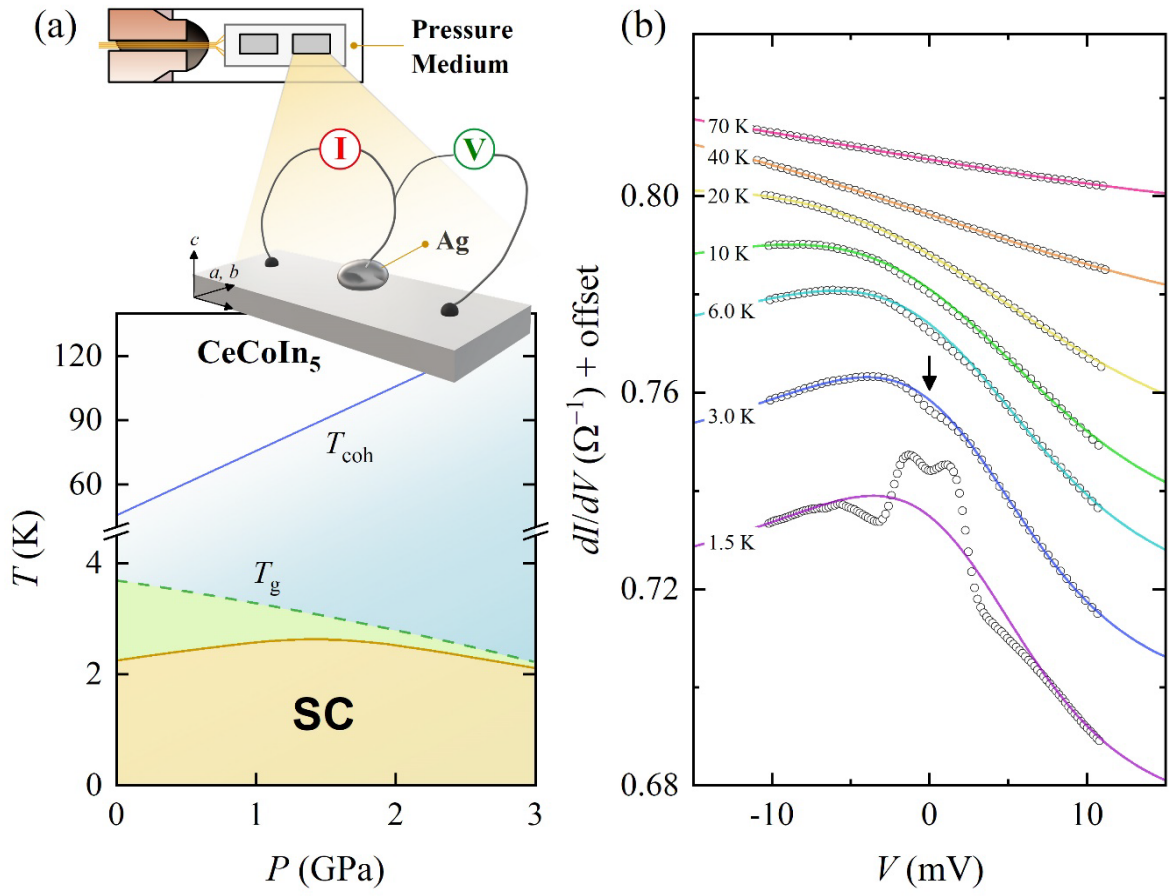

Figure 1

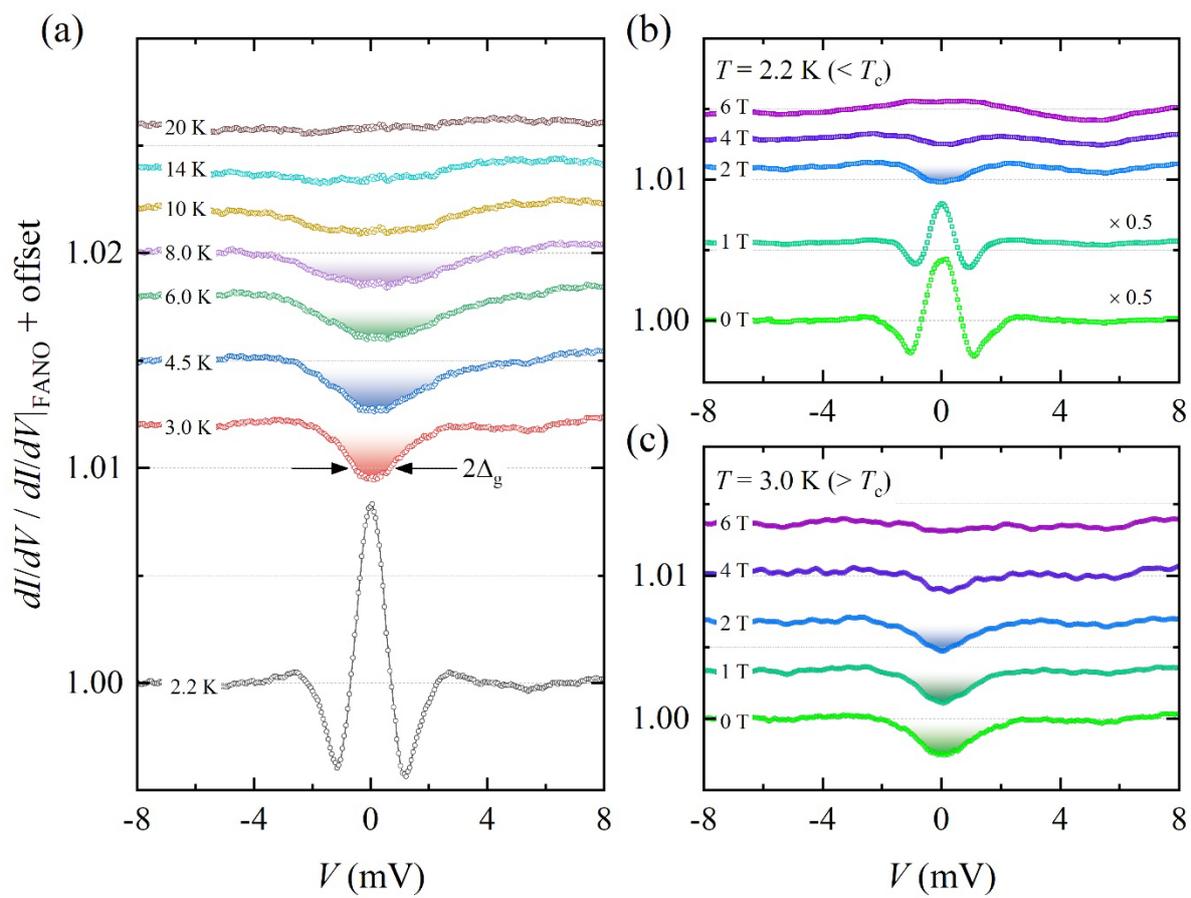

**Figure 2**

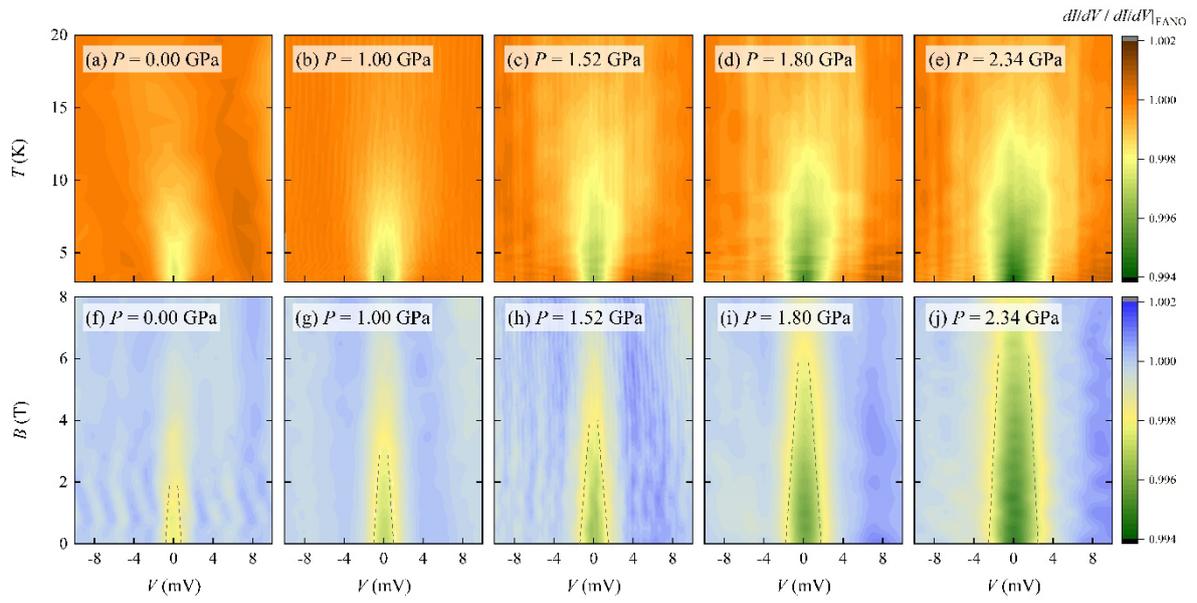

**Figure 3**

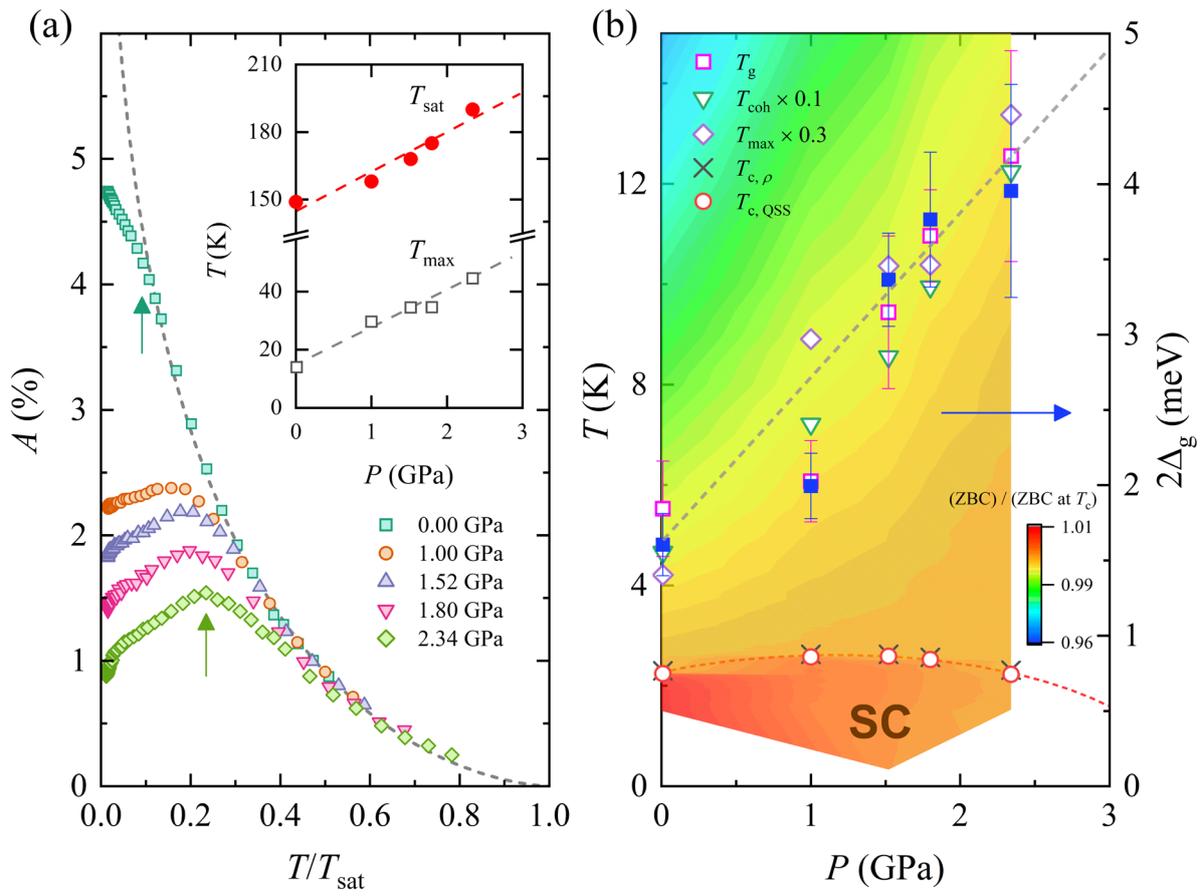

**Figure 4**

# Supplemental Material for

# Hybridization-Controlled Pseudogap State in the Quantum Critical Superconductor CeCoIn$_5$

Harim Jang[1], Vuong Thi Anh Hong[1], Jihyun Kim[1], Xin Lu[2], and Tuson Park[1,3,*]

[1] *Department of Physics, Sungkyunkwan University, Suwon 16419, Republic of Korea*

[2] *Center for Correlated Matter and Department of Physics, Zhejiang University, Hangzhou 310058, China*

[3] *Center for Quantum Materials and Superconductivity (CQMS), Sungkyunkwan University, Suwon 16419, Republic of Korea*

The SM describes additional data and analyses that support the results in the main text.

A. Experimental methods

B. Spectroscopy and its junction diagnostics

C. Fano resonance model on CeCoIn$_5$

D. Figures

E. References

## A. Experimental methods

A Pt/Ag/CeCoIn$_5$ point-contact junction is fabricated on the as-cleaved crystallographic $a$–$b$ plane of CeCoIn$_5$ single crystal to perform quasi-particle scattering spectroscopy (QSS). The small Ag powders are solutes in the organic suspension, enabling the construction of a point contact with a small effective contact area. Three different DC currents with small steps are successively applied, and the differential conductance ($dI/dV$) is obtained by calculating their slope. The average of the three measured points is used as the corresponding voltage. The moving average method is applied to minimize the effects of the electromotive force in the circuit. The small current step, typically less than 0.4 mA, is maintained over the entire process to prevent a wide energy range of scattering information at once. Electrical resistivity is measured using the conventional four-probe technique, where an AC current of 1 mA and 13.7 Hz is applied along the $a$–$b$ plane of the crystal using the Lakeshore AC resistance bridge 370.

A BeCu clamp-type piston pressure cell is used to achieve quasi-hydrostatic pressures up to 2.34 GPa and Daphne 7373 is used as a pressure-transmitting medium. The superconducting transition temperature of Pb is used as a manometer to determine the pressure inside the cell. The Cryofree Oxford TeslatronPT is used as a platform to control the temperature down to 1.5 K and a magnetic field up to 8.0 T, where the magnetic field is applied perpendicular to the crystallographic $c$ axis of CeCoIn$_5$.

## B. Spectroscopy and its junction diagnostics

The QSS is a spectroscopic tool to probe the scattering off rate of quasi-particle in condensed matter through a (metallic) point-contact junction. In the semiclassical Boltzmann's scattering theorem, the traveling electrons with the energy near the quasi-particle excitation are strongly scattered at the point-contact junction when their trajectory is close to ballistic [1]. Since the scattering rate of transmitting electrons (resistance) depends on the bias voltage, QSS has been shown to be effective to probe any kind of quasi-particles. In other words, non-Ohmic behavior of the junction as a function of electron energy serves as an energy-resolved spectroscopy, where its spectral weight represents the quasi-particle scattering rates. The QSS in the superconductor is a primary example where the Andreev reflection, the retroreflection of the hole at the junction interface owing to the formation of Cooper pair in the superconductor, takes place at the junction interface. Successful measurement and prediction of superconducting gap properties at the normal/superconductor junction has been reported for various families of superconductors, including unconventional superconductors [2-4].

Depending on the relative ratio between the point-contact junction size ($d$) and the electronic mean free path ($l$), the junction typically can be classified into three different regimes of ballistic, diffusive, and thermal [1]. In the ideal case of ballistic condition where satisfies $l \gg d$ or $l/d \gg 1$, the Fermi level is well separated as much as bias voltage $eV$ across the electrode and sample. Therefore, any kind of quasi-particle excitation energy within $eV$ can be mapped out in a spectroscopic signal. On the other hand, in the opposite limit case of the thermal regime, where $l \ll d$ or $l/d \ll 1$, inelastic scattering is dominant at the junction thus most energy of electron dissipates and causes an increase in temperature. In this case, the spectroscopic signal is wiped out and the $dI/dV$ curve reflects the bulk properties of the electrode and sample. The diffusive regime is in between two extreme limits, where the inelastic scattering is negligible and hence still can be served as a spectroscopic junction. The contact resistance $R_{PC}$ of the arbitrary size of homo-junction is known to follow Wexler's relation [5]:

$$R_{PC} = \frac{16\rho l}{3\pi d^2} + \beta \frac{\rho}{d} \tag{S1}$$

where $\beta \sim 1$ is size dependent term and $\rho$ is the bulk electrical resistivity. The first and second term in Wexler's relation corresponds to the Sharvin and Maxwell resistance, respectively, each represents the junction resistance of ballistic and thermal limit.

Diagnostics of point-contact junctions in extreme conditions of high pressure and low temperature is not straightforward because direct access to the junctions is highly limited, especially under high pressures where the junctions are surrounded by the pressure-transmitting-medium. In this study, the ratio $l/d$ is estimated based on a few assumptions: 1) The junction resistance can be written as a sum of Sharvin and Maxwell terms, thus satisfying the Wexler relation (Eq. S1). 2) the multiply of bulk electrical resistivity by elastic mean free path is constant, i.e., $\rho l \sim 10^{-11}$ $\Omega\cdot cm^2$. This value is not only valid for the elemental metal but also for the heavy-fermion system where the effective mass of the itinerant electron is increased [6-8]. This is because both Fermi velocity and heavy quasi-particle relaxation rate are renormalized by the ratio of effective mass, canceling out each other [6]. 3) The heterojunction, multichannel, and, anisotropic electrical resistivity effects are ignored.

Figure S1 shows the pressure dependence of the $R_{PC}$ and estimated $d$ and ratio $l/d$, suggesting that $l/d$ retained above 1 for all the pressure ranges. The increase of $l/d$ with increasing pressure may be owed to the increase of $l$ as the system is away from the magnetic fluctuation.

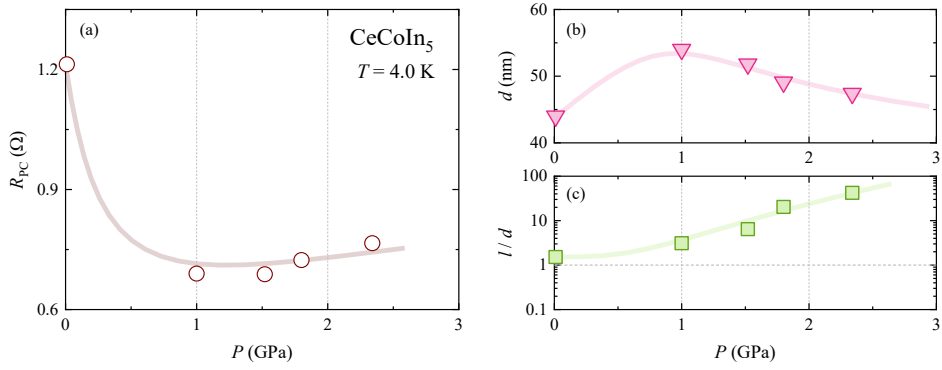

**FIG. S1. Contact resistance, junction size and ratio $l/d$ of CeCoIn$_5$ as a function of external pressure.** (a) The contact resistance ($R_{PC}$) of the Pt/Ag/CeCoIn$_5$ point-contact junction is displayed as a function of pressure, where $R_{PC}$ is determined from the zero-bias voltage limit at 4.0 K. Pressure dependence of estimated (b) junction size ($d$) and (c) ratio $l/d$ by solving Eq. S1. Junction size ranged from ~44 to ~54 nm for all the pressure ranges and $l/d$ increased nearly exponentially with increasing pressure up to 2.34 GPa. The solid lines are the guide to the eye, and the horizontal dashed line in panel (c) indicates the $l = d$ line.

In the thermal regime, the temperature of junction attributed to the dissipation is given by $T_{PC}^2 = T_0^2 + V^2/4L$, where $T_0$, $V$, and $L$ indicate the temperature of the bath, bias voltage, and Lorenz number, respectively [1,9]. Therefore, the qualitative comparison between the transformed $dI/dV$ into the temperature basis and temperature dependence of $\rho$ can instruct whether the junction at least falls in the thermal regime or not. The Lorenz number of CeCoIn$_5$ at ambient pressure is close to $L_0 = (\pi e k_B)^2/3$ and weakly depends on temperature below 20 K [10]. We assume that the Wiedemann-Franz law is valid and the Lorenz number is invariant under pressures used in this study.

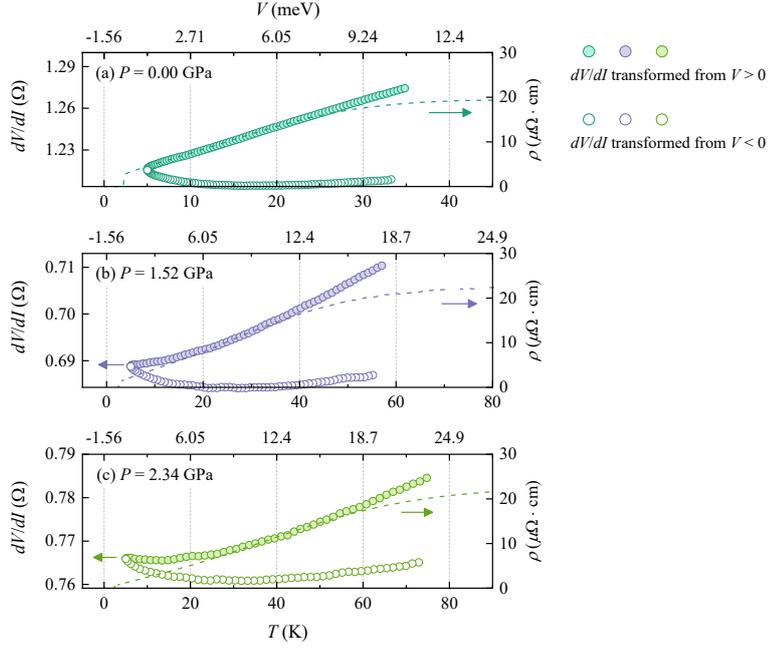

**FIG. S2. Comparison of the transformed $dV/dI$ of CeCoIn$_5$ with its bulk resistivity.** Transformed $dV/dI$ (closed and open symbols for $V > 0$ and $V < 0$, respectively) spectrum from the $dI/dV$ at $T_0 = 5.0$ K into the temperature space at representative pressures of (a) 0.00 GPa, (b) 1.52 GPa, and (c) 2.34 GPa, under the assumption that the junction belongs to the thermal regime. The relation of $T_{PC}^2 = T_0^2 + V^2/4L$ is used to estimate the junction temperature, where the Lorenz number is assumed to be constant (see above). The top $x$-axis of each panel is the bias voltage corresponding in the bottom $x$-axis. The bulk electrical resistivity ($\rho$) of CeCoIn$_5$ at corresponding pressure is plotted on the left ordinate for comparison (dashed lines). A large discrepancy at high temperatures (high bias voltages) between two quantities indicates that the junction is parted from the thermal regime.

Figure. S2 shows the comparison of the transformed $dV/dI(T_{PC})$ curves for both the positive and negative bias voltage regime at $T = 5.0$ K with the bulk $\rho(T)$. All the scales in Fig. S2 are adjusted to maximize the overlap between two curves of $dV/dI(T_{PC}, V > 0)$ and $\rho(T)$ as much as possible, for a clear comparison (see the left and right axis for the $dV/dI$ and $\rho$, respectively).

Significant gaps are found for both the low and high bias voltage regimes, and the curvature of $dV/dI(T_{PC})$ and $\rho(T)$ are entirely different for all the temperature ranges thus irreducible to each other (except for $10 < T < 25$ K at ambient pressure where both curves are linear, which may be the consequence of the coincidence). These diagnostics demonstrate that the point-contact junction does not belong to the thermal regime and may exhibit negligible inelastic scattering for the probed range.

## C. Fano resonance model in CeCoIn$_5$

Fano resonance model successfully described the experimental result of the excited states of helium under consideration of the quantum interference of the single discrete state and continuum, where the transition rate to the discrete state in the presence of the interference with the continuum is as follows [11]:

$$R(V) = R_0(V) \frac{(q+\tilde{E})^2}{1+\tilde{E}^2}, \qquad \tilde{E} = \frac{V-\varepsilon_0}{\Gamma} \tag{S2}$$

Here, $R_0(V)$ is the rate of transition without interference, and $q$ is the dimensionless Fano parameter that determines the coupling ratio of the discrete state and continuum. $\varepsilon_0$ is the position of the resonance energy level, and $2\Gamma$ is the FWHM of the resonance. The concept of Fano interference has been successfully applied to condensed matter systems containing magnetic impurities, such as the Kondo impurity system, which has a narrow width of the resonance peak in spectral density near the Fermi level [12,13]. In the metallic Kondo system, two distinct channels can be developed for transmitting electrons near the magnetic impurity site. One is the direct flow to the conduction band, which acts as a continuum channel. Coupling to the spin of magnetic impurities can be considered as a discrete state in the framework of the Anderson impurity model, thus providing an additional channel [14].

In order to describe the differential conductance of the Kondo lattice compound CeCoIn$_5$, the modified Fano resonance model by Yang is adopted in this study [15]:

$$\frac{dI}{dV}(V,T) = g_0 + g_I(T) \int_{-\infty}^{+\infty} \frac{df(E-V,T)}{dV} \frac{|q(T)-\tilde{E}(V,T)|^2}{1+\tilde{E}(V,T)^2} dE \tag{S3}$$

Here, $V$ corresponds to the voltage bias to the point-contact junction. The term $|q(T)-\tilde{E}(V,T)|^2/1+\tilde{E}(V,T)^2$ represents the Fano line as proposed in Eq. S2, where $\tilde{E}(V,T) = (E-\varepsilon)/[(a\cdot V)^2+\gamma(T)^2]^{1/2}$ is a phenomenologically modified expression. The Fano parameter $q = q_1 + iq_2$ is the tunneling ratio between two paths, where $q_2$ is the direct tunneling term into the $c$–$f$ coupled state with energy $\varepsilon$ without Fano interference. $q_2$ gives the Lorentzian contribution to $dI/dV$ owing to the strong electronic correlations. $f(E,T)$ is the Fermi function, $g_0$ is the background conductance assumed to be independent of temperature, and $\gamma(T)$ is the scattering rate at the zero-bias voltage within the discrete channel. Parameter $a$, which is introduced to describe the asymmetric behavior at a high bias voltage, is fixed to 0.5 for all fittings [15].

No additional background curve except $g_0$ is considered in the analysis because 1) $q_2$ already includes the Lorentzian shape contribution centered around $\varepsilon$, and 2) the pressure dependence of $g_0$ is consistent with the contact resistance $R_{PC}$ without extra consideration (see Fig. S1). To consider the thermal broadening of the Fermi function, integration at each bias voltage is performed with a width of $30 \times k_B T$. The best fitting results are obtained using the least-squares method from $dI/dV$, which excludes the data range showing the pseudogap feature and superconductivity. The Kondo resonance energy $\varepsilon$, background conductance $g_0$, and Fano parameter $q_1$ are fixed as constants over the entire temperature range at each pressure (see Fig. S7).

# D. Figures

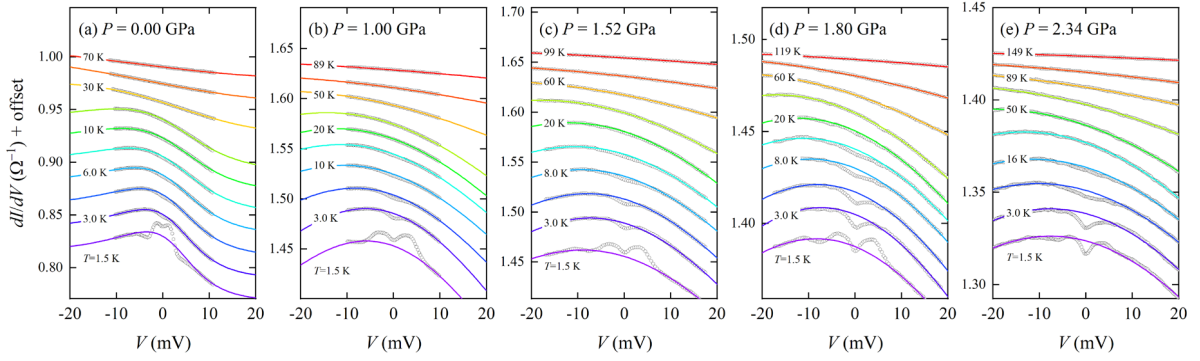

**FIG. S3. Temperature dependence of *dI/dV* of CeCoIn$_5$ at various pressures.** The differential conductance (*dI/dV*) curves at representative temperatures are selectively plotted as a function of the bias voltage from 1.5 K to the highest measured temperature at (a) 0.00 GPa, (b) 1.00 GPa, (c) 1.52 GPa, (d) 1.80 GPa, and (e) 2.34 GPa. The solid lines indicate the best fits of the Fano resonance model at each temperature. All the curves are plotted with an offset for clarity. At ambient pressure, as shown in (a), a broad hump in *dI/dV* with asymmetry appears at low temperatures, which could be ascribed to Fano resonance. With increasing temperature, the signature for Fano resonance is suppressed, leading to a small energy-dependent conductance background at sufficiently high temperatures. At temperatures below the superconducting transition temperature $T_c$, a sharp enhancement of *dI/dV* with two-peak structures arises from the Andreev reflection.

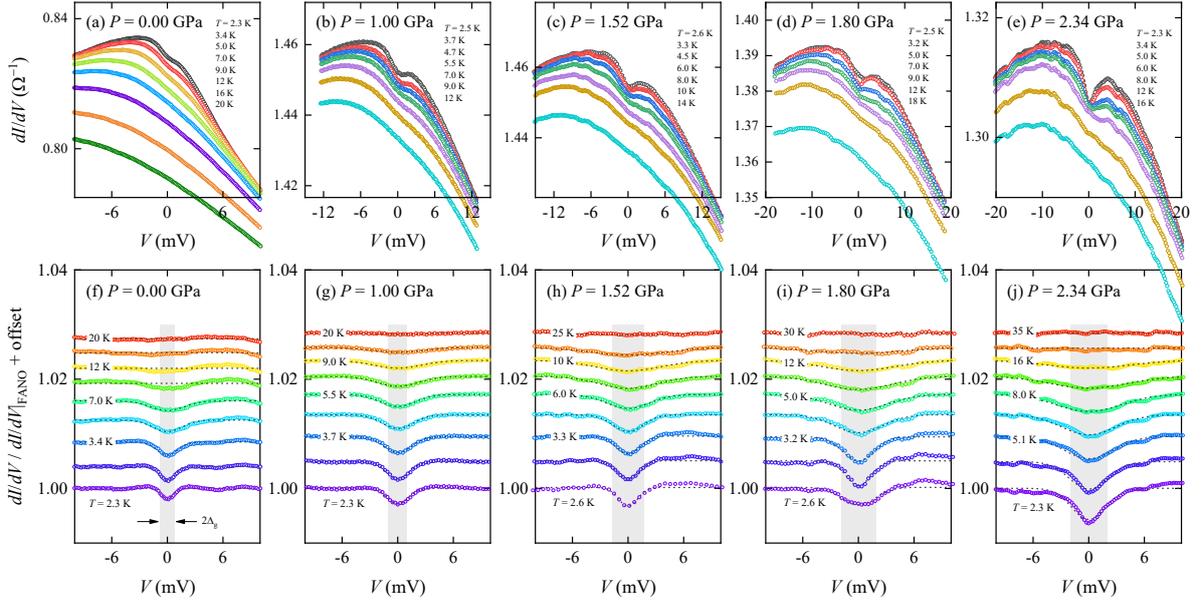

**FIG. S4. Temperature dependence of *dI/dV* and the normalized *dI/dV* at various pressures.** The raw spectrum of differential conductance *dI/dV* at selective temperatures across $T_g$ is displayed under the pressure of (a) 0.00 GPa, (b) 1.00 GPa, (c) 1.52 GPa, (d) 1.80 GPa, and (e) 2.34 GPa. The normalized *dI/dV* divided by $dI/dV|_{FANO}$, $dI/dV / dI/dV|_{FANO}$, is plotted against the bias voltage at various pressures of (f) 0.00 GPa, (g) 1.00 GPa, (h) 1.52 GPa, (i) 1.80 GPa, and (j) 2.34 GPa. Here, $dI/dV|_{FANO}$ is estimated from the Fano resonance model and each curve is plotted with an offset for clarity. The pseudogap feature is gradually smeared out with increasing temperature and finally disappears, and the *dI/dV* curve is well explained by the Fano resonance model. Suppression in the normalized conductance is fitted by the Gaussian distribution function (dashed lines), and the shaded area indicates the estimated energy width of the pseudogap $2\Delta_g$. $T_g$ is defined using the criterion of –0.2 % boundary in $dI/dV / dI/dV|_{FANO}$. We note that this method does not represent the 'real' onset of the pseudogap feature properly, as depicted here and also in Fig. S5. However, the rough criteria could capture the proper pressure dependency of the pseudogap because it is beyond the error of Fano fittings and experimental resolution at high temperatures, which is typically in the range of 0.05 %.

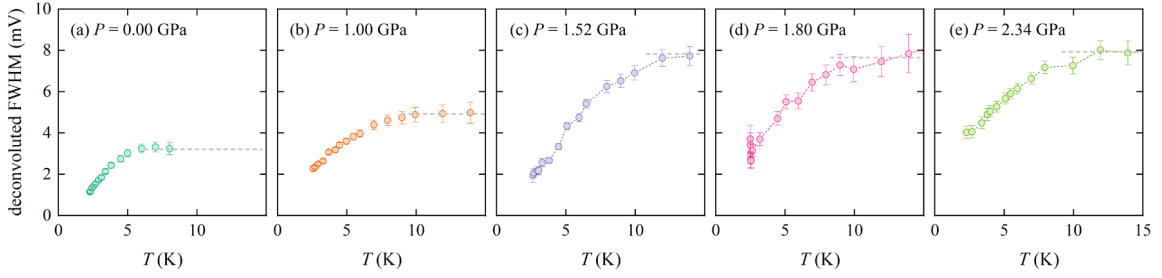

**FIG. S5. Temperature-dependence of FWHM of the pseudogap at various pressures.** The full width at half maximum (FWHM) of the pseudogap extracted by subtracting the roughly estimated thermal broadening effect ($3/2 \times k_B T$) from the total FWHM in $dI/dV / dI/dV|_{FANO}$ is plotted as a function of temperature at (a) 0.00 GPa, (b) 1.00 GPa, (c) 1.52 GPa, (d) 1.80 GPa, and (e) 2.34 GPa [16]. The deconvoluted FWHM after subtracting the thermal effect is 1.15($\pm$0.08) meV at 2.3 K and ambient pressure, which increases with increasing temperature and is saturated to approximately 3.2 meV above 6.1 K. Application of the hydrostatic pressure increases the temperature-dependent deconvoluted FWHM. At 2.34 GPa, the deconvoluted FWHM is 4.06($\pm$0.3) meV at 2.6 K and is saturated to 8.0 meV at above 11.8 K. The dashed lines are a guide to the eye to the saturated value of FWHM. The error bar is the standard deviation of the least-squares fitting of the Gaussian function.

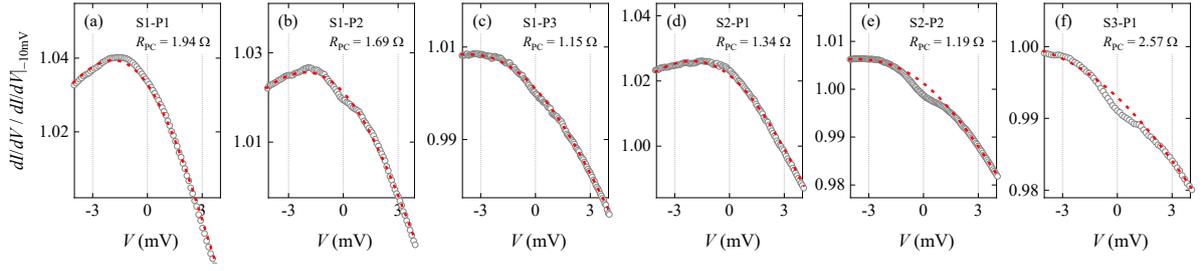

**FIG. S6. Reproducibility of pseudogap feature** (a) – (f) $dI/dV$ divided by the value at –10 mV, $dI/dV \,/\, dI/dV|_{-10\text{mV}}$, for junction S$n$-P$m$ at $T = 2.5$ K and ambient pressure, where $n$ and $m$ indicate the sample and junction number respectively. The dashed line at each panel is the best fitting result of the Fano resonance model for the given $dI/dV$ curves. Clearer pseudogap behavior can be observed for S1-P2, S1-P3, S2-P2, and S3-P1. The junction resistance $R_{\text{PC}}$ is given together at the upper right corner, implying that the $R_{\text{PC}}$ is less correlated to the pseudogap behavior. In this work, the junction S2-P2 is mainly used. The appearance of pseudogap behavior depends on the junction, which may be ascribed to the sensitive pseudogap to the surface state the junction faced. This has been pointed out by several atomic-scale studies on the surface of CeCoIn$_5$ using STS, revealing that the pseudogap only manifested in the specific layer of crystal that shows a strong coupling to $f$ electrons [16,17]. This underlines that the pseudogap behavior is correlated quantum phenomena with the Kondo hybridization. Therefore, the point-contact junction may not always display the pseudogap behavior in the normal state, and it would depend on which crystal surface state is dominant in the junction. Searching for the origin of surface dependence of pseudogap in CeCoIn$_5$ together with microscopic understanding will be an important future work.

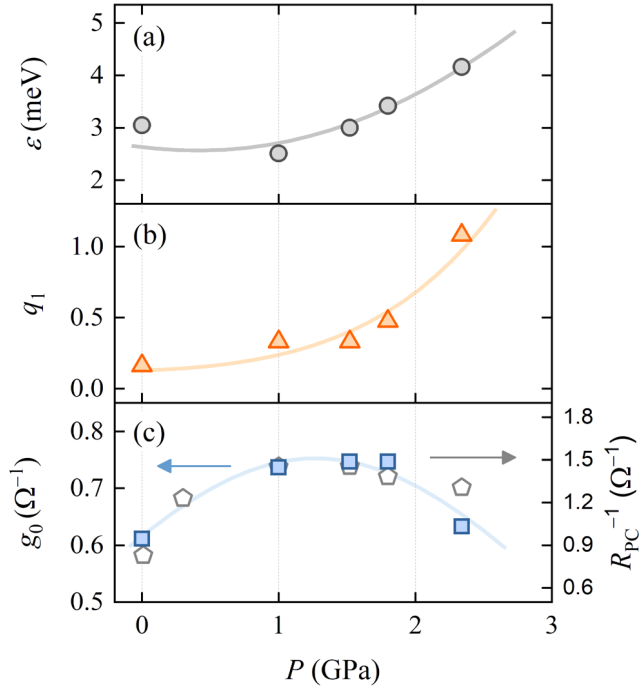

**FIG. S7. Pressure-dependence of the fitting parameters obtained from the Fano resonance model.** (a) Kondo resonance energy $\varepsilon$, (b) Fano parameter $q_1$, and (c) background conductance $g_0$ obtained from the Fano resonance model are plotted against pressure. The inverse of the contact resistance (right y-axis) is plotted together with the background conductance (left y-axis) in (c) for comparison. The Kondo resonance energy $\varepsilon$ is 3.05 meV at ambient pressure and gradually increases with pressure, reaching 4.16 meV at 2.34 GPa. The Fano parameter $q_1$ also increases from 0.16 at ambient pressure to 1.08 at 2.34 GPa, indicating that the external pressure shifts the Kondo resonance energy higher and increases the c–f coupling rate of transmitting electrons. As shown in (c), the pressure-dependent background conductance shows a similar tendency with the contact resistance at 4.0 K, suggesting a correlation between the two properties. The solid lines are guides to the eye.

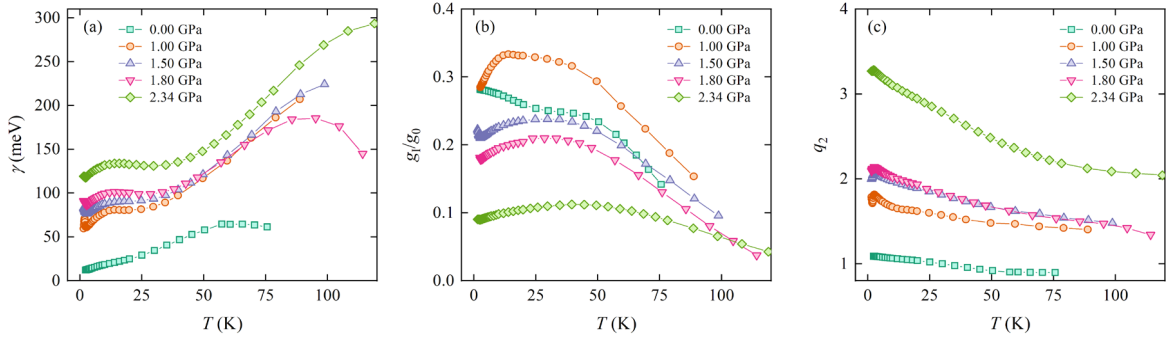

**FIG. S8. Pressure- and temperature-dependence of the Fano fitting parameters $\gamma$, $g_I/g_0$, and $q_2$ at various pressures for CeCoIn$_5$.** The temperature-dependence of Fano fitting parameters (a) $\gamma$, (b) $g_I/g_0$, and (c) $q_2$ at various pressures. The zero-bias scattering rate $\gamma$, which is associated with the energy width of the Fano resonance, increases as temperature increases at all pressures. The estimated zero-temperature limit of $\gamma(T=0)$, $\gamma_0$, from a simple extrapolated quadratic line, is approximately 10.6 meV at ambient pressure and increases steeply to 112 meV at 2.34 GPa. In the Kondo impurity case, the resonance width attributed to the Kondo exchange is $\gamma_0 = k_B T_K$, where $T_K$ is the Kondo temperature [13]. The estimation of $T_K$ is approximately 120 K at ambient pressure, but 1300 K at 2.34 GPa, which seems too high at high pressure, indicating that additional effects may be required to describe $dI/dV$ in Kondo lattice CeCoIn$_5$ under pressure. The ratio of the two conductance terms $g_I$ and $g_0$, $g_I/g_0$, initially increases with temperature, but decreases at high temperatures, showing a maximum in its temperature-dependence as with the degree of the asymmetry in the $dI/dV$ spectrum (see the main text). The imaginary part of the Fano parameter, $q_2$, is extrapolated to 1.09 at 0 K, which decreases with increasing temperature. As pressure increases, the zero-temperature limit value of $q_2$ increases monotonically up to 3.35 at 2.34 GPa.

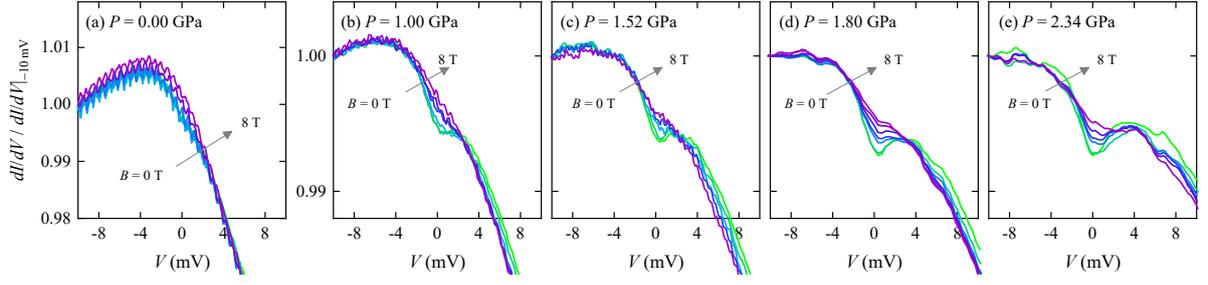

**FIG. S9. Magnetic field-dependence of pseudogap feature at 3.0 K.** The magnetic field-dependence of $dI/dV$ divided by the value at $-10$ mV, $dI/dV / dI/dV|_{-10\text{mV}}$, is plotted against the bias voltage for various magnetic fields at (a) 0.00 GPa, (b) 1.00 GPa, (c) 1.52 GPa, (d) 1.80 GPa, and (e) 2.34 GPa. The green-colored line indicates data at 0 T, whereas the purple line is for data at 8 T. Here, the magnetic field is applied perpendicular to the crystallographic $c$ axis. Application of the magnetic field suppresses the pseudogap feature, but does not alter the differential conductance value at the high bias region outside the energy gap. With increasing pressure, the critical field to suppress the pseudogap becomes higher: 8.0 T is insufficient at a pressure of 2.34 GPa.

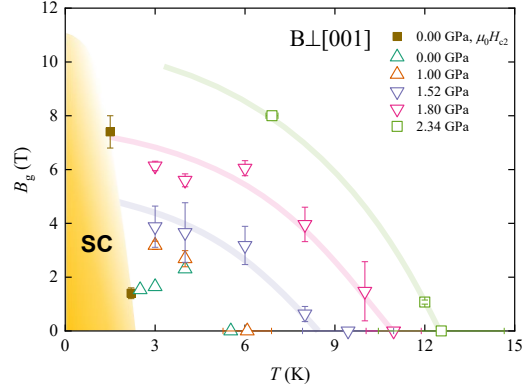

**FIG. S10. *T–B* phase diagram of the pseudogap.** The critical magnetic field, $B_g(T)$, is plotted as a function of temperature at various pressures. Here, the critical magnetic field $B_g$ that suppress the pseudogap feature is defined by the same criteria with characteristic temperature, $T_g$, at which $dI/dV$ deviates from the Fano resonance model by $-0.2\%$. The closed symbol shows the temperature-dependent upper critical fields at ambient pressure and the colored area indicates the superconducting state. Application of the pressure increases both $B_g$ and $T_g$, enlarging the area of the pseudogap state in the *T–B* phase diagram. The solid lines are guides to the eye.

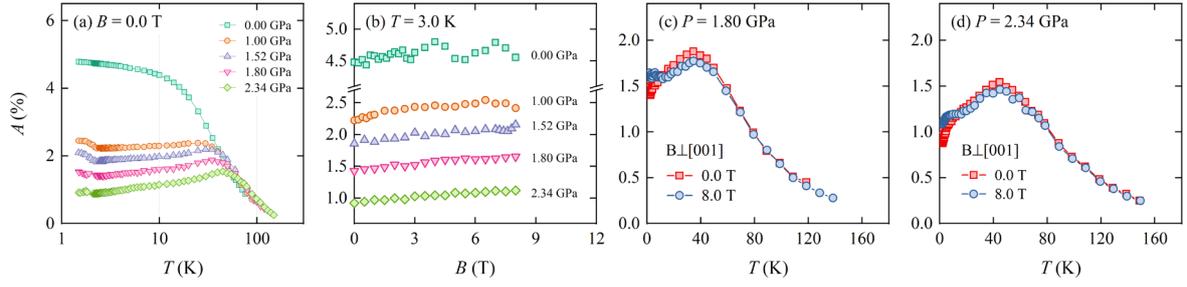

**FIG. S11. Effect of magnetic field on the asymmetry in differential conductance spectrum.** (a) Asymmetry $A$ (%) = $100 \times (dI/dV|_{-10\text{mV}} - dI/dV|_{+10\text{mV}}) / (dI/dV|_{+10\text{mV}})$ at 0 T is plotted as a function of temperature on a semi-logarithmic scale at various pressures. (b) The magnetic field dependence of $A$ is plotted at 3.0 K and various pressures, which is almost independent of the magnetic field up to 8.0 T. Temperature-dependence of $A$ is comparatively plotted between 0.0 T and 8.0 T at (c) 1.80 GPa and (d) 2.34 GPa, respectively. With decreasing temperature, the asymmetry increases logarithmically and changes its behavior near the peak temperature $T_{\text{max}}$. In contrast, it is almost independent of the magnetic field up to 8.0 T for all pressures. The lack of change in the asymmetry under magnetic field may be ascribed to the smaller energy scale of the Zeeman energy ($\mu_B B \approx 0.46$ meV) than the Kondo resonance width $\gamma_0$ ($\sim k_B T_K \approx 10$ meV) estimated from the single-site Kondo resonance limit.

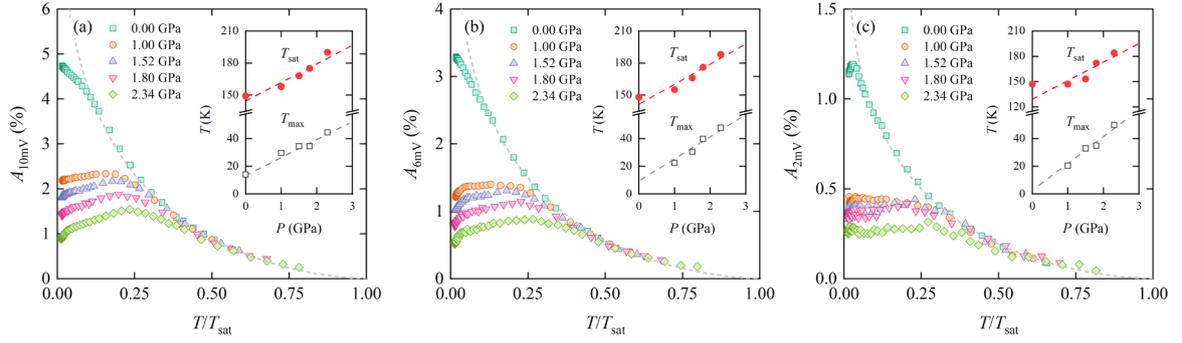

**FIG. S12. Asymmetry in differential conductance spectrum from various definitions.** In general, asymmetry in this study can be defined by $A_X$ (%) = 100 × $(dI/dV|_{-X} - dI/dV|_{+X}) / (dI/dV|_{+X})$, where X is the bias voltage to the junction in the unit of mV. The asymmetry in $dI/dV(V)$ using the criterion of (a) X = 10 mV, (b) X = 6 mV, and (c) X = 2 mV is plotted as a function of the reduced temperature ($T/T_{sat}$) at various pressures, where $T_{sat}$ is the onset temperature where the asymmetry starts to appear. The dashed line is the asymmetry in the two-fluid model of the Kondo lattice system (see the main text). Each inset shows the pressure-dependence of $T_{sat}$ and $T_{max}$ determined from each definition, where $T_{max}$ is the temperature where the asymmetry shows a maximum or kink. Even though the magnitudes of asymmetry decrease with decreasing X, all $A_X(T)$ shows the main features as follows: (i) the maximum (or kink) at low temperature, (ii) collapse to a single curve in the high-temperature regime, and (iii) the similar linear pressure-dependence and values of $T_{sat}$ and $T_{max}$. These results show that the selection of X that is used to determine asymmetry does not affect our analysis.

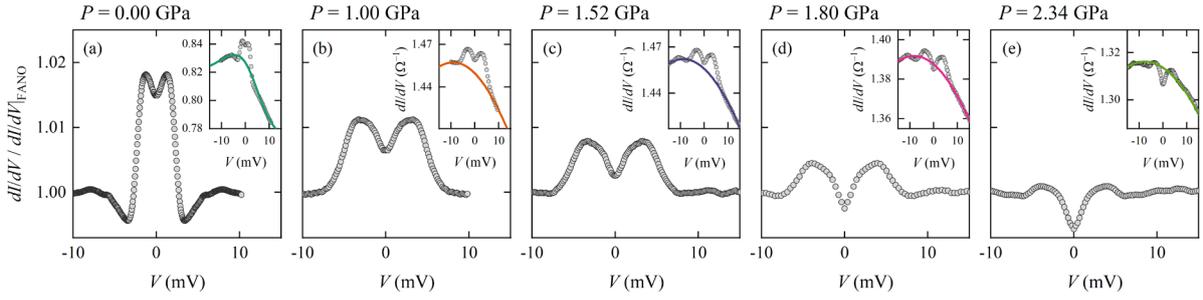

**FIG. S13. Differential conductance curves at 1.5 K.** $dI/dV$ normalized by the calculated Fano-line background ($dI/dV|_{\text{FANO}}$) at 1.5 K is representatively shown for (a) 0.00 GPa, (b) 1.00 GPa, (c) 1.52 GPa, (d) 1.80 GPa, and (e) 2.34 GPa. The solid lines in the inset are the Fano background at each pressure, which is calculated from the conductance curve after excluding the Andreev reflection signal. All the curves are symmetrized by averaging the $dI/dV$ of positive and negative bias sides, after the normalization process for clarity. The two-peak structure in the $dI/dV$ of CeCoIn$_5$ at 1.5 K is the hallmark of the Andreev reflection in the superconducting state. Under pressure conditions, all the $dI/dV \,/\, dI/dV|_{\text{FANO}}$ curves show a similar Andreev reflection structure, indicating that our Pt/Ag/CeCoIn$_5$ junction has not been significantly altered by the pressurizing process. The size of the superconducting energy gap ($\Delta_{\text{SC}}$) estimated by the peak-to-peak distance ($=\Delta_{\text{SC, PP}}$) is ~1.2 meV at 1.5 K and ambient pressure. This value is slightly larger than $\Delta_{\text{SC}}$ of ~0.6 to ~0.9 meV obtained from STS and ~0.5 meV in QSS, indicating that a quasi-particle lifetime parameter $\Gamma$ may need to be considered, owing to the multichannel effects in the soft-point contact junction [3,18,19]. Under pressure, $\Delta_{\text{SC, PP}}$ shows a significant increase to 3 meV at 1.0 GPa, reaches a maximum near 1.80 GPa, and slightly decreases at 2.34 GPa, similar to the behavior of $T_c$ against pressure. The fact that a similar broadening effect is also observed in the conductance asymmetry $A$ (see the main text) in the pressure range where the contact resistance $R_{\text{PC}}$ decreases supports the multichannel effect in the junction.

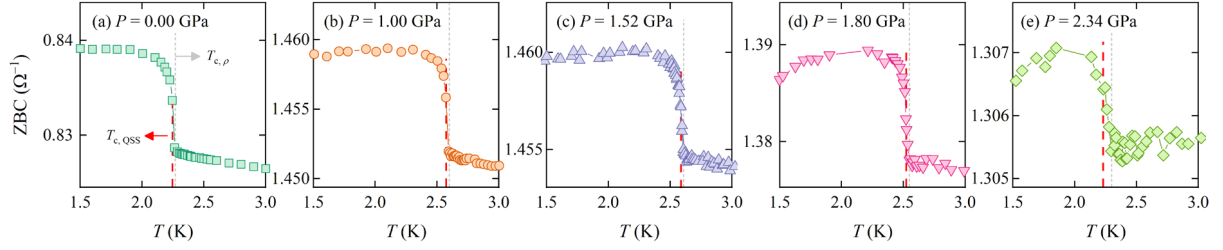

**FIG. S14. Temperature-dependence of the zero-bias-conductance at various pressures.** The zero-bias conductance (ZBC), the differential conductance value at the zero-bias voltage, is plotted as a function of temperature for different pressures. The dotted vertical lines correspond to $T_{c,\rho}$, which is defined as the mid-point of superconducting transition in a temperature-dependent electrical resistivity drop. The dashed vertical lines indicate $T_{c,\text{QSS}}$, a mid-point of ZBC drop. The consistency of two transition temperatures suggests that the heating effect may not important in the temperature range studied in this work.

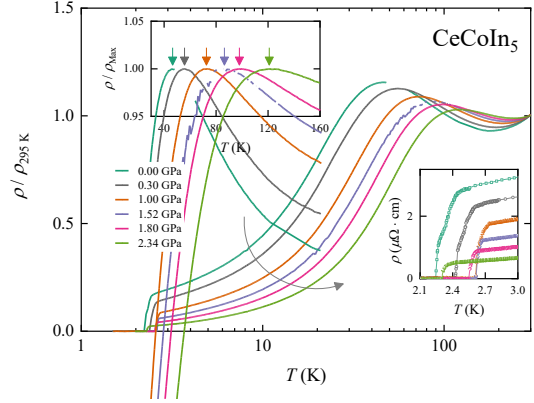

**FIG. S15. Electrical resistivity of CeCoIn₅ as a function of temperature at various pressures.** The temperature-dependence of the electrical resistance of CeCoIn₅ divided by the value at 295 K, i.e., $\rho / \rho_{295K}$, is displayed on a semi-logarithmic scale at representative pressures. The upper inset depicts the normalized resistivity to the transport Kondo coherence temperature ($T_{coh}$), whereas the lower inset is the resistivity near the superconducting transition temperature. The resistance increases logarithmically as the temperature lowers and shows a maximum at the coherence temperature, $T_{coh}$. Application of the pressure promotes the collective excitations of c–f coupling, leading to an increase in $T_{coh}$: $T_{coh}$ of 45 K at ambient pressure increases monotonically to approximately 122 K at 2.34 GPa. The superconducting transition temperature $T_{c,\rho}$, which is defined from the mid-point of resistance drop, is 2.3 K at ambient pressure and shows the dome shape centered around 1.52 GPa (see Fig. 1 in the main text).

# E. References